\documentclass[aps,prl,superscriptaddress,showpacs,twocolumn]{revtex4}
\usepackage{units}
\usepackage{amsmath}
\usepackage{amssymb}
\usepackage{graphicx}
\usepackage{bm}
\usepackage{multirow,color}

\newcommand{\be}{\begin{equation}}
\newcommand{\ee}{\end{equation}}
\newcommand{\e}{\epsilon}
\newcommand{\bj}[2]{J_{#1}(#2)}
\newcommand{\bh}[2]{H_{#1}(#2)}
\newcommand{\bjp}[2]{J'_{#1}(#2)}
\newcommand{\bhp}[2]{H'_{#1}(#2)}

\newcommand{\bhs}[1]{H_{#1}}

\newcommand{\bhps}[1]{H'_{#1}}

\newcommand{\cs}[1]{\cos(#1\theta)}
\newcommand{\re}[1]{\text{Re}[#1]}

\begin{document}
\title{Extreme sensitivity of output directionality to boundary perturbation in wavelength-scale microcavities}

\author{Li Ge}
\affiliation{Department of Electrical Engineering, Princeton University, Princeton, NJ 08544, USA}
\author{Qinghai Song}
\affiliation{Department of Electronic and Information Engineering, Shenzhen Graduate School, Harbin Institute of Technology, Shenzhen, 518055, China}
\author{Brandon Redding}
\affiliation{Department of Applied Physics, Yale University, New Haven, CT 06520-8482, USA}
\author{Hui Cao}
\affiliation{Department of Applied Physics, Yale University, New Haven, CT 06520-8482, USA}

\date{\today}

\begin{abstract}
We report a surprising observation that the output directionality from wavelength-scale optical microcavities displays extreme sensitivity to deformations of the cavity shape. A variation of the cavity boundary on the order of ten thousandth of a wavelength may flip the output directions by 180 degrees.  Our analysis based on a perturbation theory reveals that a tiny shape variation can cause a strong mixing of nearly degenerate cavity resonances with different angular momenta, and their interference determines the farfield emission pattern. This work shows the possibility of utilizing carefully-designed wavelength-scale microcavities for high-resolution detection and sensing applications.
\end{abstract}

\maketitle

Optical microcavities have a wide range of applications from lasers, filters, sensors to cavity quantum electrodynamics and single-photon emitters \cite{Vahala}. The cavity shape has been used as a design parameter to control the output coupling \cite{MekisPRL95, NockelNat97, GmachlSci98, NarimanovPRL99, LeePRL02, ChernAPL03, BaryshnikovPRL04, SchwefelJOSAB04, LebentalAPL06, GaoAPL07, TanakaPRL07, WiersigPRL08, WangPNAS10}, and a key issue is the sensitivity to small variations. In a billiard with closed boundary condition, the effect of a tiny change of the cavity shape accumulates as light undergoes specular reflections inside the cavity. For example, the intracavity ray dynamics changes from regular to chaotic as the cavity evolves from a circle to a stadium shape \cite{Bunimovich}. However, light in a dielectric cavity has a finite lifetime before escaping, and the sensitivity to the cavity boundary variation is thus reduced. Nevertheless, dramatically different emission patterns were observed from similarly deformed cylindrical polymer lasers \cite{SchwefelJOSAB04}. The observation was explained by the distinct geometries of the unstable manifolds, which determine the refractive escape routes at large deformation.
For a smaller deformation from a circle or sphere, the output is dominated by tunneling instead of refraction \cite{NockelNat97,LaceyPRL03, PodolskiyOL05, CreaghPRL07, ShinoharaPRL10,CreaghPRE12}, and the evanescent field becomes highly directional even with a weak deformation. Experiments using nearly spherical resonators \cite{LaceyPRL03, XiaoOL09} attributed the observed directional tunneling to nonperturbative phase space structures in the intracavity ray dynamics.

All these studies were performed in the semiclassical regime, where the cavity size $R$ is much larger than the wavelength $\lambda$. As such, the variation of the boundary, though small compared to $R$, is comparable to or even larger than the wavelength. It would be interesting to see what happens in the wave regime where $R$ approaches $\lambda$ \cite{SongPRL10, SongPRA11, ReddingPRL12}. A variation of the boundary with similar size relative to $R$ is then on a scale much smaller than the wavelength. Can light resolve such a minute structural feature, and if so in what way?

To answer these questions, we preformed numerical studies on deformed wavelength-scale dielectric microdisks. Our results show that variations of the cavity boundary on the scale of $\lambda/10^{4}$ not only can switch the output of a cavity mode from bidirectional to unidirectional, but can also flip the output directions by 180 degrees. To understand such high sensitivity, we applied a perturbation theory to the wave problem \cite{SI}. The shape variation causes a mixing of modes with different angular momenta, and their interference outside the cavity determines the farfield emission pattern. If two modes have nearly degenerate frequencies, they can be strongly mixed, and the degree of their mixing is extremely sensitive to the perturbation of the cavity boundary. Such extreme sensitivity brings the opportunity of utilizing carefully-designed wavelength-scale microcavities for high-resolution detection and sensing applications.

In the discussion below we characterize the boundary of a microdisk cavity by $\rho(\theta)=R[1+ \e_2 \cs{2} + \e_3 \cs{3}]$ in the polar coordinates, where $|\e_2|,\, |\e_3| \ll 1$. For a small deformation the dipolar term ($\e_1 \cos\theta$) mostly leads to a lateral shift of the cavity, and it is eliminated by choosing a proper origin of the coordinate system. For a wavelength-scale cavity $R < \lambda$, where $\lambda$ is the wavelength in vacuum, or equivalently, $k R < 2 \pi$ where the wavevector $k = 2 \pi / \lambda$. Below we discuss transverse electric (TE) modes (electric field parallel to the disk plane) which are most common in microdisk lasers, and the same effects also exist for transverse magnetic (TM) modes.

\begin{figure}[h]
\centering
\includegraphics[width=\linewidth]{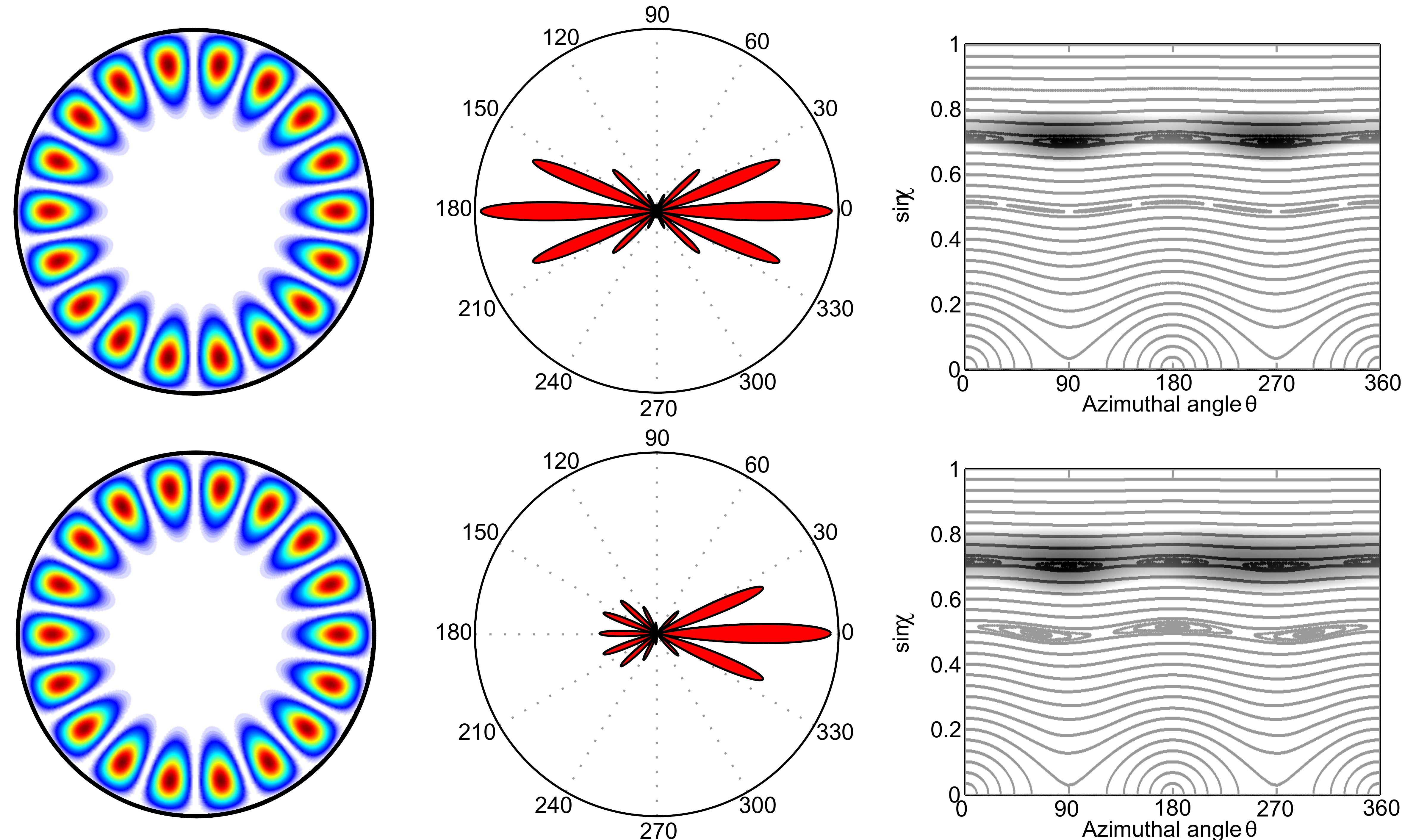}
\caption{(Color online) Top row: Internal structure (left), farfield pattern (middle), and Husimi Projection on the SOS (right) of Mode 1 at $kR=4.387-i1.809\times10^{-5}$ in a quadrupole cavity with $R=1\,\text{$\mu$m}$, $\e_2=-0.01$, $\e_3=0$, and $n=3$. The black solid contour represents the farfield obtained from the 2nd order perturbation theory, which agrees almost exactly with the numerical data (red shadow). The classical ray dynamics in a billiard of the same shape but with closed boundary is also shown on the SOS. The six islands around $\sin\chi=0.5$ correspond to the forward (``$\vartriangleright$") and backward (``$\vartriangleleft$'') triangular orbits, and the four islands near $\sin\chi=0.7$ correspond to the diamond orbit (``$\diamond$"). Bottom row: Same as in the top row but the cavity is now sightly perturbed with $\e_3 =10^{-4}$. The resonance shifts slightly to $kR=4.387-i2.039\times10^{-5}$. The classical SOS is mostly unchanged, except that the islands of the forward triangular orbit disappear.} \label{fig:e3}
\end{figure}

We first consider a series of slightly deformed quadrupolar cavities with refractive index $n=3$, $R=1\,\text{$\mu$m}$, $\e_2=-0.01$, and varying $\e_3$ with $|\e_3|<10^{-3}$. Using a scattering matrix approach \cite{NarimanovPRL99,HP1} we calculated the cavity resonant frequencies and quality ($Q$) factors. Some of the  high-$Q$ modes exhibit dramatic output sensitivity to the value of $\e_3$. Take the resonance at $\re{kR}=4.387$ (Mode 1) as an example, its output is bidirectional towards $\theta=0^\circ,\,180^\circ$ at $\e_3=0$ (see the Supplemental Material \cite{SI}); but when $\e_3$ becomes $10^{-4}$, the emission of Mode 1 is greatly suppressed along $\theta=180^\circ$, giving rise to an enhanced forward ($\theta=0^\circ$) emitting pattern (Fig.~\ref{fig:e3}). By flipping the sign of $\e_3$, the output direction of Mode 1 is reversed, since now the cavity becomes the mirror image of the previous one about the vertical axis, i.e. $\rho(\pi-\theta)=R[1+\e_2\cs{2}-\e_3\cs{3}]$. Despite the drastic change of the farfield pattern, the spatial pattern of the resonance inside the cavity barely changes as shown in Fig.~\ref{fig:e3}.

The observed extreme boundary sensitivity cannot be explained by semiclassical ray dynamics \cite{NockelNat97,GmachlSci98}, which can be mapped onto the Poincar\'e Surface of Section (SOS) using the positions of rays incident on the boundary (represented by the azimuthal angle $\theta$) and the corresponding angles of incidence $\chi$. As shown in Fig.~\ref{fig:e3}, the majority of the SOS remains regular in the presence of small $\e_2$ and $\e_3$, with unbroken Kolmogorov-Arnold-Moser (KAM) curves transversing the entire cavity boundary ($\theta\in[0, 360^{\circ}]$). There are a few islands corresponding to stable periodic orbits. When $\e_3$ changes from 0 to $10^{-4}$, the islands at $\sin \chi \sim 0.5$ display a noticeable change. However, the high-$Q$ resonances stay at significantly higher $\sin \chi$, where the SOS remains almost the same. For example, Fig.~\ref{fig:e3} shows the Husimi projection \cite{HP1,HP2} of Mode 1, from which we see that it is localized at $\sin \chi \sim 0.7$ and has little overlap with those islands at $\sin \chi \sim 0.5$. Such high-$Q$ modes can be considered as quasi-whispering gallery (WG) modes, and their output comes from direct tunneling to the leaky region where $\sin \chi < 1 / n$. The tunneling rate is the largest where the cavity boundary has the highest curvature, i.e. at $\theta=90^\circ,\,270^\circ$ for $\e_3=0$. These positions only shift by about $1^\circ$ when $\e_3=10^{-4}$, which cannot account for the strong asymmetry in the farfield pattern.

\begin{figure}[b]
\centering
\includegraphics[width=\linewidth]{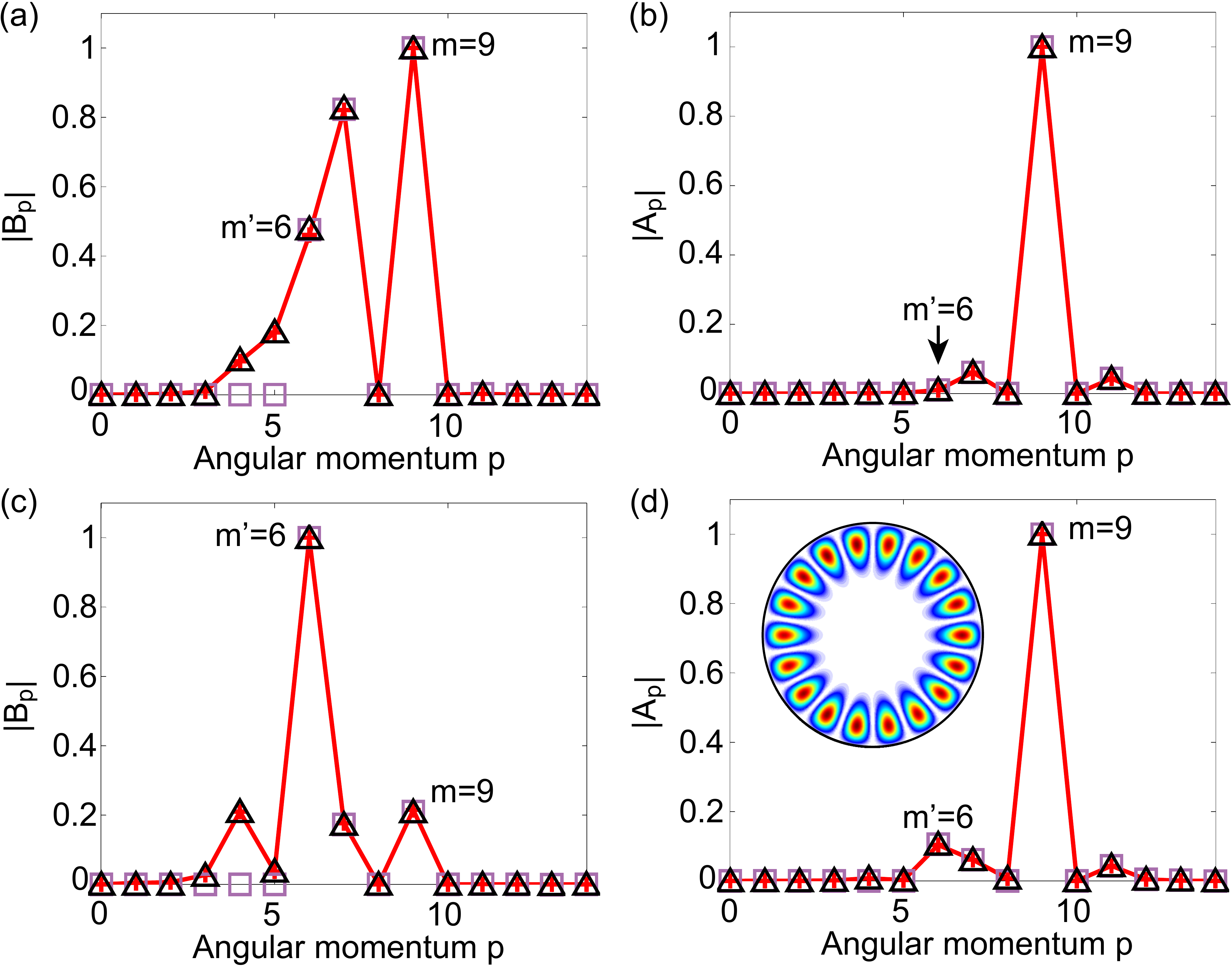}
\caption{(Color online) Amplitude of normalized Hankel (a,\,c) and Bessel (b,\,d) coefficients outside and inside the cavity (red crosses connected by solid line) with $R=1\,\text{$\mu$m}$, $n=3$, $\e_2=-0.01$, $\e_3=10^{-4}$ (a,\,b), $10^{-3}$ (c,\,d). Purple squares and black triangles are given by the 1st and 2nd order perturbation calculation, respectively.  Although the dominant angular momentum outside has changed from $m=9$ to $m'=6$ as $\e_3$ increases from $10^{-4}$ to $10^{-3}$, inside the cavity the $m=9$ component is still the dominant one. Inset in (d) shows the mode structure inside the cavity at $\e_3=10^{-3}$, which is almost identical to that shown in Fig.~\ref{fig:e3} at $\e_3=10^{-4}$.}\label{fig:e3_coeff}
\end{figure}

To gain insight into the extreme sensitivity of the output on the boundary deformation, we turn to the angular momentum analysis of the modal wavefunction inside and outside the cavity,
\be
\psi(r,\theta) =
\begin{cases}
 \sum_p A_p \bj{p}{nkr}\cos(p\theta), & r<\rho(\theta), \\
 \sum_p B_p \bh{p}{kr}\cos(p\theta), & r>\rho(\theta).
\end{cases}\label{eq:ansatz}
\ee
$\bj{p}{nkr}$, $\bh{p}{kr}$ are the $p$-th order Bessel function and the outgoing Hankel function, respectively.  $A_p$ ($B_p$) will be referred to below as the Bessel (Hankel) coefficients inside (outside) the cavity. Because the cavity has reflection symmetry with respect to the horizontal axis, the cavity resonances have either even parity or odd parity about $\theta=0$. Here we consider the even modes, which can be decomposed by $\cos(p \theta)$. The analysis of the odd modes is similar with $\cs{p}$ replaced by $\sin(p\theta)$. Since $|\e_2|,\, |\e_3| \ll 1$, a quasi-WG mode has a dominant angular momentum component $m$ inside the cavity. For Mode 1, $m=9$ [Fig.~\ref{fig:e3_coeff}(b, d)]. The quadrupolar deformation $\e_2\cs{2}$ scatters light from $m$ to $m\pm2$. Since the $m+2$ component is confined within the cavity more strongly, $m$ and $m-2$ components are dominant outside the cavity at $\e_3=0$. They interfere destructively in the $\theta = 90^{\circ}, 270^{\circ}$ directions, and constructively in the $\theta = 0^{\circ}, 180^{\circ}$ directions, giving rise to the bidirectional output \cite{SI}. As $\e_3$ becomes nonzero, the $\cos(3 \theta)$ deformation introduces additional $m \pm 3$ components, with $m-3$ stronger than $m+3$ outside the cavity. Consequently, the dominant Hankel coefficients are $m, m-2, m-3$ as shown in Fig.~\ref{fig:e3_coeff}(a) at $\e_3 = 10^{-4}$; they not only have comparable amplitudes but also similar phases. Since $\cs{6}$ is symmetric about the vertical axis while $\cs{7}$ and $\cs{9}$ are antisymmetric, it interferes negatively with the other two in the $\theta = 180^{\circ}$ direction while interfering positively in the $\theta = 0^{\circ}$ direction, causing the unidirectional emission shown in Fig.~\ref{fig:e3}.

\begin{figure}[h]
\centering
\includegraphics[width=\linewidth]{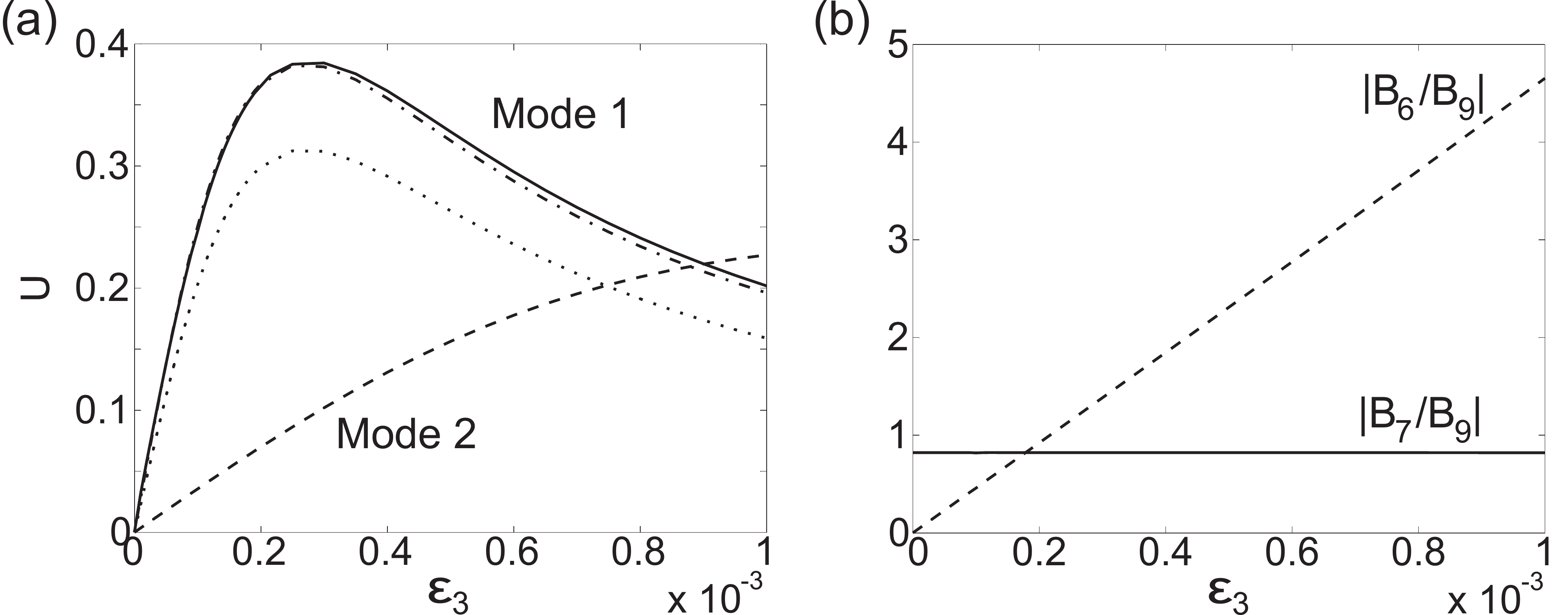}
\caption{(a) $U$ versus $\e_3$ for Mode 1 (solid line) and 2 (dashed line) in Fig.~\ref{fig:quasiDeg}(a). 1st order (dotted line) and 2nd order (dash-dotted line) perturbation results for Mode 1 are also shown. (b) Ratio of Hankel coefficients $|B_7/B_9|$ (solid line) and $|B_6/B_9|$ (dashed line) in Mode 1 as a function of $\e_3$. Parameters used are the same as the bottom row in Fig.~\ref{fig:e3}.}\label{fig:sensitivity}
\end{figure}

In the meanwhile, the corresponding Bessel coefficient $A_{6}$ remains much smaller than $A_{9}$ [Fig.~\ref{fig:e3_coeff}(b)], thus it barely alters the mode structure inside the cavity. This holds true when $\e_3$ increases to $10^{-3}$, at which $B_{6}$ dominates over $B_9$ and $B_7$ outside the cavity. The increasing amplitude difference of even and odd Hankel waves reduces the interference effect, bringing down the unidirectionality, which can be measured by $U \equiv \int_0^{2\pi} d\theta I(\theta)\cos\theta$ from the normalized farfield intensity $I(\theta)$. $U$ is zero for isotropic or bi-directional emission, and positive (negative) for unidirectional emission in the forward (backward) direction. Fig.~\ref{fig:sensitivity}(a) shows $U$ as a function of $\e_3$. As $\e_3$ increases, $U$ of Mode 1 rapidly increases to its maximum of $0.39$ at $\e_3 \simeq 2.7 \times 10^{-4}$ before it decreases gradually.

\begin{figure}[t]
\centering
\includegraphics[width=\linewidth]{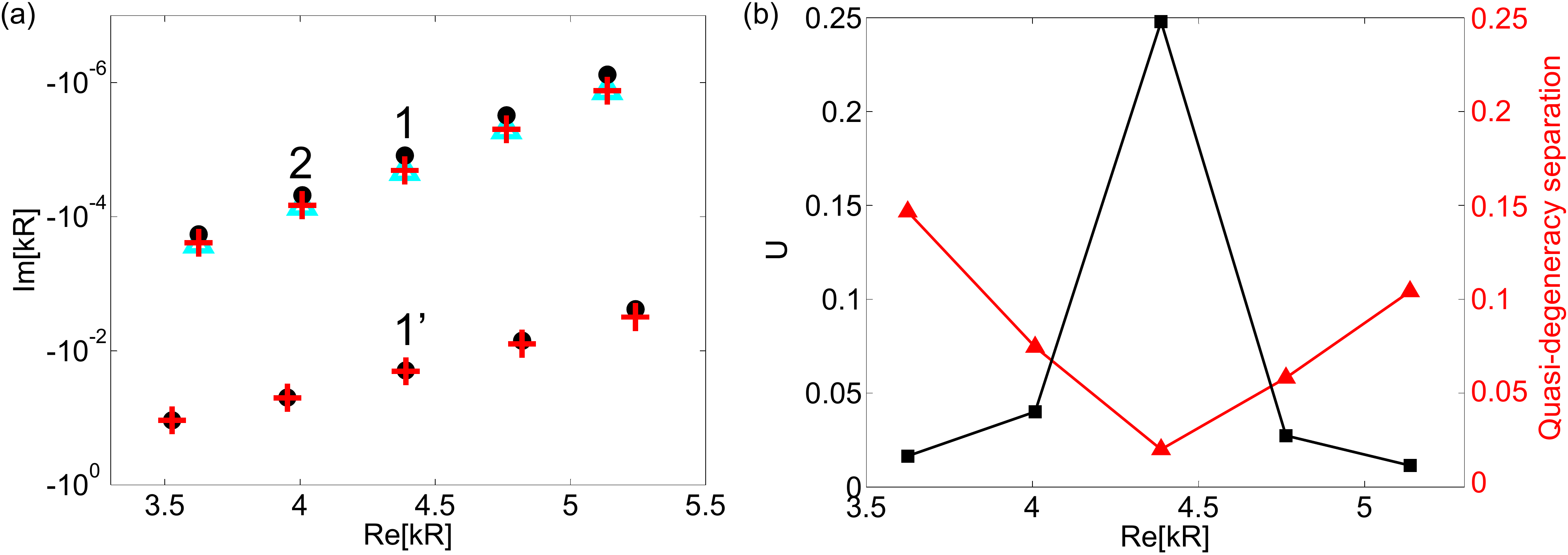}
\caption{(Color online) (a) Complex resonance frequency $kR$ (red crosses) in a cavity with $R=1\,\text{$\mu$m}$, $\e_2=-0.01$, $\e_3=10^{-4}$, and $n=3$. The corresponding resonances in a circular disk of the same $R$ are marked by black dots. Cyan triangles are given by the 2nd order perturbation theory. (b) Unidirectionality $U$ of the high-$Q$ resonances (black squares) and their distances to the nearest low-$Q$ resonances in the complex frequency plane (red triangles) versus $\re{kR}$ of the high-$Q$ modes. }\label{fig:quasiDeg}
\end{figure}

The analysis above illustrates that the change of output directionality is a coherent wave effect. What is surprising is the rapid growth of $B_6$ with $\e_3$. Fig.~\ref{fig:sensitivity}(b) shows the ratio $|B_6/B_9|$ and $|B_7/B_9|$ as a function of $\e_3$. $B_6$ already surpasses $B_9$ in amplitude at $\e_3$ as small as $2.2\times10^{-4}$. How can such a tiny perturbation [$\e_3 \cs{3}$] of cavity boundary cause a strong mixing of even and odd angular momenta? To identify the cause, we noticed that there is a lower-$Q$ resonance whose frequency is close to Mode 1. Fig.~\ref{fig:quasiDeg}(a) shows a series of high-$Q$ modes with nearly constant frequency spacing, and a low-$Q$ series with slightly larger spacing. Fig.~\ref{fig:quasiDeg}(b) reveals a correlation between unidirectionality of a high-$Q$ mode and its spacing to the nearby low-$Q$ mode. At $\e_3 = 10^{-4}$, Mode 1 has the largest $U$ among all the high-$Q$ resonances in Fig.~\ref{fig:quasiDeg}(a), and its distance to its quasi-degenerate partner (Mode $1'$) is also the shortest. Mode $1'$ has a dominant angular momentum $m'=6$ \cite{SI}, which appears in Mode 1 at $\e_3 \neq 0$. These results suggests a coupling between Mode 1 and $1'$. Recent studies \cite{WiersigPRA06, SongPRL10} demonstrate that a high-$Q$ mode can acquire unidirectional emission from a low-$Q$ mode to which it couples. However, this scenario does not happen here, because Mode $1'$ emits more or less symmetrically in the forward and backward directions as shown in Ref.~\cite{SI}. The nature of mode interaction here is completely different.

To gain physical insight about the extreme boundary sensitivity, we adopted a perturbation theory \cite{NgJOSAB02, DubertrandPRA08, LeePRL08} to the TE modes in a slightly deformed microdisk cavity, $\rho(\theta)=R+\e f(\theta)$, where the dimensionless $\e\ll 1$ \cite{SI}. Since the cavity is nearly circular and the resonances under consideration are quasi-WG modes, we chose the resonances $k_0$ of a circular cavity of radius $R$ as our unperturbed basis and treated the deformation $\e_2 \cs{2}+ \e_3\cs{3} = \e f(\theta)$ as the perturbation. $k_0$ is determined by the boundary condition for TE modes in a circular cavity, which gives $T_m(k_0R)\equiv (1/n) \bjp{m}{nk_0R}/\bj{m}{nk_0R}- \bhp{m}{k_0R}/\bh{m}{k_0R}=0$. In the deformed cavity, the resonance frequency can be expanded as $k=k_0+k_1\e + k_2\e^2 + O(\e^3)$. For convenience, we write $A_p = {a_p}/{\bj{p}{nkR}}$, $B_p = (a_p+b_p)/{\bh{p}{kR}}$ in (\ref{eq:ansatz}) and normalize $\psi(\vec{r})$ by scaling the dominant $a_m$ to unity. In Ref.~\cite{SI} we show that all $a_{p\neq m}$ and $b_p$ are at least of order $\e^1$, thus we define $a_{p\neq m} \equiv \alpha_p\e + \beta_p\e^2 + O(\e^3)$ and $b_p \equiv \mu_p\e + \gamma_p\e^2 + O(\e^3)$. By expanding the TE boundary conditions \cite{SI} to $\e^2$ around $r=R,\,k=k_0$, we find the corrections to the resonance frequency $k$ as well as the coefficients $a_p$ and $b_p$.

With the second order corrections $\beta_{p\neq m}$ and $\gamma_{p}$ given in Ref.~\cite{SI}, the perturbation theory reproduces the numerical results nicely (see Figs.~\ref{fig:e3}-\ref{fig:quasiDeg}). In fact, the essence of the extreme boundary sensitivity of the output directionality is already well captured by the first order corrections
\begin{gather}
\alpha_p = \frac{1}{T_p}\left[ k_0RS_m\left(\frac{\bhps{p}}{\bhs{p}}-\frac{\bhps{m}}{\bhs{m}}\right)-T'_m\right]F_{pm},\;({p\neq m})\label{eq:alpha0}\\
\mu_p = k_0RS_m F_{pm},\label{eq:mu0}
\end{gather}
as shown in Fig.~\ref{fig:sensitivity}(a). We have defined $S_p(x)\equiv n \bjp{p}{nx} / \bj{p}{nx} - \bhp{p}{x} / \bh{p}{x}$, $F_{pm} \equiv {c_p} \int_0^{2\pi} f(\theta)\cs{p}\cs{m} d\theta/{2\pi R}\, (c_{p}=2-\delta_{p,0})$ and dropped the arguments of the Bessel and Hankel functions and their derivatives.
We note that the first order correction to the resonance, $k_1 = -\e k_0 F_{mm}$, vanishes unless $f(\theta)$ changes the average radius (i.e., $\int f(\theta) d \theta \neq 0$), thus the second order treatment is needed to capture the shift of the resonances (see Fig.~\ref{fig:quasiDeg}(a)). The main sidebands $m\pm2$ at $\e_3=0$ are not affected by the variation of $\e_3$, since they do not couple to each other or $m$ via $\e_3$. This holds true both inside and outside [Fig.~\ref{fig:sensitivity}(b)] the cavity.

The presence of another resonance $k_0'R$ with a dominant angular momentum $m'$ in close vicinity of $k_0R$ implies that $T_{m'}(k_0R)\approx T_{m'}(k'_0R)=0$. When this occurs, the $m'$ component in $\psi^{(m)}(\vec{r};k_0)$ is much enhanced via $\alpha_{m'}$, since $T_{m'}^{-1}(k_0R)\gg1$. This large prefactor amplifies the small boundary variation of $\cos{(m-m')\theta}$, especially when the $m'$ component is leakier ($m'<m$) and has a stronger influence on the wavefunction outside the cavity. For example, the unperturbed WG resonance corresponding to Mode 1 is $k_0R=4.388-i1.226\times10^{-5}$ with $m=9$, and there is a quasi-degenerate resonance of lower-Q at $k'_0R=4.391-i1.153\times10^{-2}$ with $m'=6$. The factor $|T_{m'}^{-1}(k_0R)|=7.930$ is much larger than its typical value in the absence of quasi-degeneracy. As a result, $\alpha_{m'}$ increases rapidly with $F_{mm'}=\e_3/2$, so does $B_{m'}$ with respect to $B_{m}$. Although $\alpha_{m'}$ appears in the Bessel coefficient $A_{m'}$ as well, $|B_{m'}/B_{m}|$
increases much more rapidly than $|A_{m'}/A_{m}|$ because the scale factor $|\bh{m}{kR}/\bh{m'}{kR}|$ is much larger than $|\bj{m}{nkR}/\bj{m'}{nkR}|$. In Mode 1 the latter is almost 50 times smaller than the former, which explains the almost identical mode structure inside the cavity while the farfield pattern changes dramatically with $\e_3$.

Another important factor for the extreme sensitivity is the phase of $\alpha_{m'}$, which differs from $a_{m}(=1)$ by $\pi/2$ as given by (\ref{eq:alpha0}). With an extra $\pi/2$ relative phase coming from the the asymptotic form of the Hankel function in the farfield, i.e. $H_p(kr\rightarrow\infty)\propto\exp(-ip\pi/2)$, the $m'$ component interferes constructively with the $m$ and $m-2$ components in the forward direction and destructively in the backward direction. For example, at $\e_3=10^{-4}$, the farfield intensity $I(\theta)\approx[\cs{9}+0.8\cs{7}+0.5\cs{6}]^2$.

We also checked another high $Q$ mode, Mode 2 with $m=8$ in Fig.~\ref{fig:quasiDeg}(a). Its output directionality $U$ exhibits a lower sensitivity to the change of $\e_3$ compared to Mode 1 [Figs.~\ref{fig:sensitivity}(a) and \ref{fig:quasiDeg}(b)]. It is attributed to the larger distance to its nearest low-$Q$ mode with $m'=5$, which causes less perturbation on Mode 2 ($|T_{m'}^{-1}(k_0R)|=1.924$). However, when $U$ of Mode 1 starts to decrease at larger $\e_3$, $U$ of Mode 2 keeps increasing [Fig.~\ref{fig:sensitivity}(a)] and reaches a maxima of 0.237 at $\e_3 \simeq 1.3\times10^{-3}$ (not shown).

In summary, we show that the emission of high-$Q$ resonances in wavelength-scale microdisk cavities can be changed dramatically when the cavity boundary is modified on a scale much smaller than the wavelength ($\sim\lambda/10^4$). Such extreme sensitivity results from mixing of quasi-degenerate resonances, and it is expected to survive, at least partially, with small boundary roughness. Although the output directionality is affected by the boundary roughness, its sensitivity to the modification of the cavity boundary on top of a given surface roughness can still be observed in the numerical simulation \cite{SI}, as long as quasi-degenerate modes exist in the presence of boundary roughness. Our findings may have applications in high-resolution detection and sensing applications, for example, in detecting and controlling thermal vibrations when combined with optomechanics techniques \cite{Vahala_Science07}.

We thank Eugene Bogolmony for bringing to our attention the boundary perturbation theory for slightly deformed microdisk cavities (Ref.~\cite{DubertrandPRA08}). We also acknowledge Remy Dubertrand, Jan Wiersig, Alex Ebersp\"acher, and Hakan T\"ureci for helpful discussions. Q.~S. acknowledges partial support by the open project of the State Key Laboratory on Integrated Optoelectronics (No. 2011KFB005). B.~R. and H.~C. acknowledge NSF under the Grants No. ECCS-1068642 and ECCS-1128542.

\end{document}